\documentclass[12pt]{elsart}
\usepackage{graphics}

\begin{document}
\begin{frontmatter}

\title{A Fiber Bundle Model of Traffic Jams}

\author{Bikas K. Chakrabarti}\ead{bikask.chakrabarti@saha.ac.in}
\address{Theoretical Condensed Matter Physics Division and\\Centre for 
Applied Mathematics and Computational Science,\\Saha Institute of 
Nuclear Physics, 
1/AF Bidhan Nagar, Kolkata 700064, India.}


\begin{abstract}
We apply the equal load-sharing fiber bundle model
of fracture failure in composite materials to model the traffic failure in a
system of parallel road network in a city. For some special distributions of
traffic handling capacities (thresholds) of the roads, the critical behavior
of the jamming transition can be studied analytically. This has been compared 
with that for the asymmetric simple exclusion process in a single channel
or road.
\end{abstract}

\end{frontmatter}

\section {Introduction}

Traffic jams or congestions occur essentially due to the excluded volume 
effects (among the vehicles) in a single  road and due to the 
cooperative (traffic) load sharing by the (free) lanes or roads in
multiply connected road networks
(see e.g., \cite{Chowdhury:2000}). We will discuss here
briefly how the Fiber Bundle Model (FBM) for fracture failure in composite
materials (see e.g, \cite{Pradhan:2003}) can be easily adopted for
the study of 
(global) traffic jam in 
a city network of roads. In the equal-load-sharing FBM, the load carried 
by the failed fibers is transferred and shared uniformly by the other
surviving fibers in the bundle and therefore the dynamics of failure
propagation in the system is analytically tractable \cite{Pradhan:2003}.
Using this
model for the traffic network, it is shown here that the generic
equation for the approach of the jamming transition in FBM corresponds to
that for the Asymmetric Simple Exclusion Processes (ASEP) leading 
to the transport failure transition in a single channel or lane (see
e.g., \cite{Stinchcombe:2001}).

\section {Model}
Let the suburban highway traffic, while entering the city, get
fragmented equally through the various  narrower streets within the 
city and get combined again outside the city (see Fig. 1). Let 
the input traffic current 
be denoted by $I_O$ and the total output traffic current be denoted by $I_T$.
In the steady state, without any global traffic jam, $I_T  =  I_O$. In
case, $I_T$ falls below $I_O$, the global jam starts and soon $I_T$ drops to
zero. This occurs if $I_O  >  I_c$, the traffic current in the network beyond 
which global traffic jam occurs. The steady state flow picture then looks 
as shown in Fig. 2. Let us assume that the parallel roads within the city 
have got different thresholds for traffic handling capacity: 
$i_{c_1}, i_{c_2}, \ldots, i_{c_N}$ for the $N$ different roads 
(the $n$-th road gets jammed if the 
traffic current $i$ per road exceeds $i_{c_n}$). Initially $i = I_O/N$ and
increases as some of the roads get jammed and the same traffic load $I_i$
has to be shared equally by a lower number of unjammed roads. Next, we
assume that the distribution $\rho(i_c)$ of these thresholds is uniform
(see Fig. 3) 
upto a maximum threshold current (corresponding to the widest road traffic
current capacity), which is normalized to unity (sets the scale for $I_c$).

\begin{figure}
\label{themodel}
\centering\resizebox*{11cm}{!}{\includegraphics{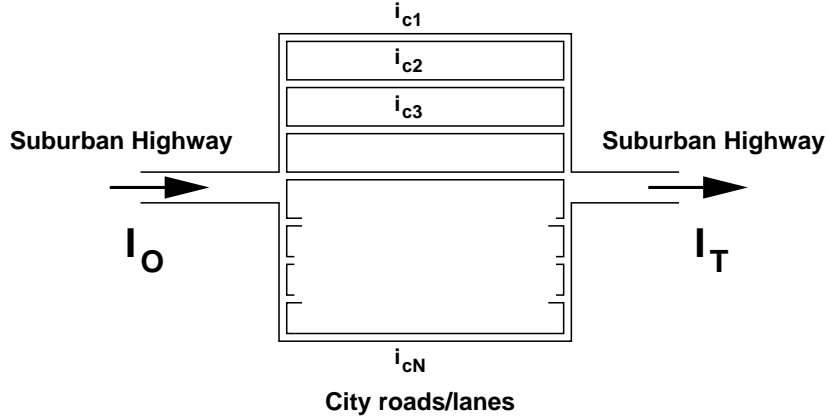}}
\caption{
The highway
traffic current $I_O$ gets fragmented into $i's$ and the narrower roads
having threshold current $i_{c_n} \le i$ gets congested or blocked and this
extra traffic load per uncongested roads get equally distributed, causing 
further blocking of some more roads.
}
\end{figure}

The jamming dynamics in this model starts from the $n$-th road 
(say in the morning)
 when the traffic load  \( i  \)
 per city roads exceeds the threshold $i_{c_n}$ of that road. 
Due to this jam, total
number of intact or uncongested roads  decreases and rest of these 
roads 
have to bear the entire traffic load in  the system. Hence effective
traffic load or stress
on the uncongested roads increases and this compels some more roads to 
fail or get jammed.
These two sequential operations, namely the stress or traffic load
 redistribution and further
failure in service of roads continue till an equilibrium is reached, where
either the surviving roads are strong (wide) enough to share equally
and  carry the
entire traffic load on the system (for $I_O < I_c$)
 or all the roads fail (for $I_O \ge I_c$)  and a (global) 
traffic jam occurs in the entire) road network system (to be brought back
perhaps in the night when the load $I_O$ decreases much below $I_c$).

\section{Jamming dynamics}

This jamming dynamics can be represented by recursion
relations in discrete time steps. 
Let us  define $U_t(i)$ to be the fraction of uncontested roads 
 in the network
that survive after (discrete) time step $t$, counted from the time $t=0$
when the load (at the level $I_O = iN$) is put in the system
 (time step indicates the number of stress 
redistributions). As such, $U_t(i =0)=1$ for all $t$ and $U_t(i)=1$
for $t=0$ for any $i$; 
$U_t(i)=U^*(i) \ne 0$ for $t \to \infty$ if 
$I_O < I_c$, and 
$U_t(i)=0$ for $t \to \infty$ if $I_O > I_c$.

\begin{figure}
\label{thejam}
\centering\resizebox*{6cm}{!}{\includegraphics{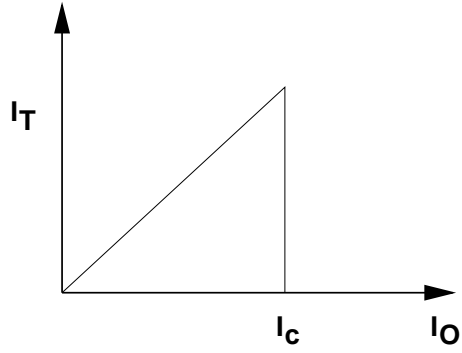}}
\caption{
Traffic jam (global) in the network of roads
occurs when $I_O~ > ~ I_c$, the threshold traffic capacity of the city
road-network.
}
\end{figure}

\begin{figure}
\label{uniform}
\centering\resizebox*{7cm}{!}{\includegraphics{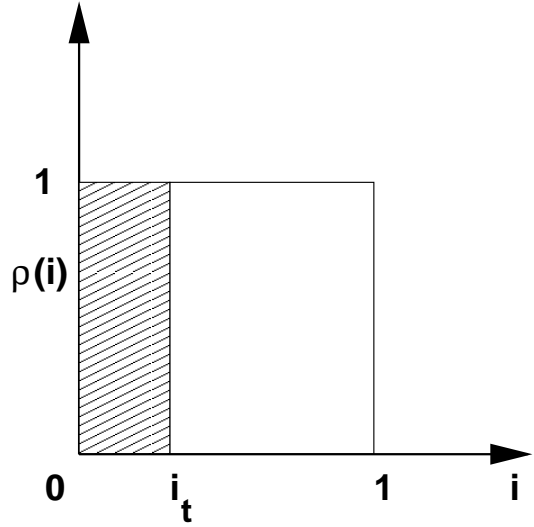}}
\caption{
The simple model considered here assumes uniform density $\rho (i)$
of the road traffic flow capacity  distribution up to a cutoff 
capacity (normalized to 
unity). At any traffic load per road level $i_t$ at time $t$, the fraction 
$i_t$ of roads fails and fraction $1-i_t$ survives.
}
\end{figure}

Therefore $U_{t}(i)$ follows a simple recursion relation (see Fig. 3) 
\[
U_{t+1}= 1-i_t;\ \ i_t = \frac{I_O}{U_t N}
\]
\begin{equation}
\label{recU_t}
{\rm or,} \ \ U_{t+1}=1-\frac{i }{U_{t}}.
\end{equation}
In equilibrium \( U_{t+1}=U_{t}=U^{*} \) and thus (1) is quadratic in
\( U^{*} \) : 
\[
U^{*^{2}}-U^{*}+i =0.
\]

\noindent The solution is

\[
U^{*}(i )=\frac{1}{2}\pm (i_{c}-i )^{1/2}; i_{c}=\frac{1}{4}.
\]

\noindent Here \( i_{c} = I_c/N \) is the critical value of 
traffic current (per road)
beyond which the bundle fails completely. 
 The quantity
\( U^{*}(i ) \) must be real valued as it has a physical meaning:
it is the fraction of the roads that remains in service under
a fixed traffic load  $i$ when the traffic load per road lies
in the range \( 0\leq i \leq i_{c} \). Clearly, \( U^{*}(0)=1 \).
 The solution
with (\( + \)) sign is therefore the only physical one.
Hence, the physical  solution can be written as 
\begin{equation}
\label{Ustarsigma_c}
U^{*}(i )=U^{*}(i_{c})+(i_{c}- i )^{1/2};
\ U^*(i_c) = \frac{1}{2} \ {\rm and}\ i_{c}=\frac{1}{4}.
\end{equation}
For \( i > i_{c} \) we can not get a real-valued fixed
point as the dynamics never stops until \( U_{t}=0 \) when the network
gets completely jammed.

\section {Critical behavior}

\noindent \textbf{(a) At \(i < i_{c} \)}
\vskip.2in
\noindent It may be noted that the quantity \( U^{*}(i )-U^{*}(i_{c}) \)
behaves like an order parameter that determines a transition from
a state of partial failure of the system(\( i \leq i_{c} \)) to a state
of total failure (\( i > i_{c} \)) :
\begin{equation}
\label{Ustar}
O\equiv U^{*}(i)-U^{*}(i_{c})=(i_{c}-i  )^{\beta };\beta =\frac{1}{2}.
\end{equation}

To study the dynamics away from criticality ($i \rightarrow i_{c}$
from below), we replace the recursion relation (\ref{recU_t}) by a differential
equation 
\[
-\frac{dU}{dt}=\frac{U^{2}-U+i }{U}.
\]

\noindent Close to the fixed point we write \( U_{t}(i )=U^{*}(i ) \)
+ \( \epsilon  \) (where \( \epsilon \rightarrow 0 \)). This
gives  
\begin{equation}
\label{epsilon}
\epsilon =U_{t}(i )-U^{*}(i )\approx \exp (-t/\tau ),
\end{equation}

\noindent where \( \tau =\frac{1}{2}\left[ \frac{1}{2}(i_{c}-i )^{-1/2}+1\right]  \).
Thus, near 
 the critical point (for jamming transition) we can write \begin{equation}
\label{dec19}
\tau \propto (i_{c}-i )^{-\alpha };\alpha =\frac{1}{2}.
\end{equation}
 Therefore the relaxation time diverges following a power-law as \( i
 \rightarrow i_{c} \)
from below.

One can also consider the breakdown susceptibility \( \chi  \), defined
as the change of \( U^{*}(i ) \) due to an infinitesimal increment
of the traffic stress \( i \) 
 \begin{equation}
\label{sawq}
\chi =\left| \frac{dU^{*}(i )}{d i }\right| 
=\frac{1}{2}(i_{c}- i )^{-\gamma };\gamma =\frac{1}{2}.
\end{equation}

\noindent  Hence the susceptibility diverges as
the applied stress \( i\) approaches the critical value \( i_{c}=\frac{1}{4} \).
Such a divergence in \( \chi  \) can be clearly observed in the
numerical study of these dynamics.

\vskip.2in
\noindent \textbf{(b) At} \textbf{\large \(i  =i_{c} \)}{\large\par}
\vskip.2in
\noindent At the critical point (\( i =i_{c} \)), we observe
a different dynamic critical behavior in the relaxation of the failure process.
From the recursion relation (\ref{recU_t}), it can be shown
that decay of the fraction \( U_{t}(i_{c}) \) of uncongested roads 
that remain in service at time \( t \) follows a simple power-law decay
:
\begin{equation}
\label{qqq}
U_{t}=\frac{1}{2}(1+\frac{1}{t+1}),
\end{equation}

\noindent starting from \( U_{0}=1 \). For large \( t \) (\( t\rightarrow \infty  \)),
this reduces to \( U_{t}-1/2\propto t^{-\delta } \); \( \delta =1 \);
a power law, and is a robust characterization of the critical
state.

\section{Universality class of the model}

The universality class of the model can be checked (cf. \cite{Pradhan:2003}) 
taking
two other types of road capacity distributions $\rho(i)$: (I) linearly 
increasing
density distribution and (II) linearly decreasing density distribution
of the current thresholds within the  limit $0$ and $1$. One can 
show that while 
$i_{c}$ changes with different strength distributions 
($i_c= \sqrt{4/27}$ for case (I) and $i_c=4/27$ for case II), 
the critical behavior remains unchanged: $\alpha =1/2=\beta =\gamma$, 
$\delta =1$ for all these equal (traffic) load sharing models.

\section {Transport in ASEP}

\begin{figure}
\label{theasep}
\centering\resizebox*{8cm}{!}{\includegraphics{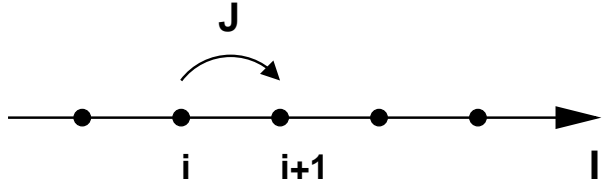}}
\caption{
 The transport current $I$ in the one
 dimensional lane or road is possible if, say, the $i$-th site is occupied and
the $(i+1)$-th site is vacant. The inter-site hopping probability is indicated
by $J$. 
}
\end{figure}

Let us consider the simplest version of the asymmetric 
simple exclusion process transport in a chain (see Fig. 4).
Here  transport corresponds to movement of vehicles, which is
possible only when a vehicle at site $i$, say, moves to the
vacant site $i+1$.
 The transport current $I$ is then given by (see e.g., 
\cite{Stinchcombe:2001}) 
\begin {equation}
\label{add1}
 I = J\rho_i (1 - \rho_{i+1}),
\end{equation}
where $\rho_i$ denotes the site occupation density at site $i$
and $J$ denotes the inter-site hopping probability. The above
equation can be easily recast in the form 
\begin{equation}
\label{add2}
 \rho_{i+1} = 1 - {\frac {\sigma} {\rho_i}}, 
\end {equation}
where $\sigma = I/J$. Formally it is the same as the recursion relation (1) for 
the density of uncongested roads in the FBM model discussed above; the
site index here in ASEP plays the role of time index in FBM.

\section {Summary and discussion}

 We have  applied here  the equal load-sharing fiber bundle model
of fracture failure in composite materials to 
model the traffic failure in a network
of parallel roads in a city. For some special distributions of
traffic handling capacities (thresholds) of the roads, namely uniform
and also linearly increasing or decreasing density of road capacities,
the critical behavior
of the jamming transition has been studied analytically. 
This dynamics of traffic jam progression (given by the recursion
relation (1), for uniform density of road capacity,
 as shown in Fig. 3 ) has been compared
with that for the asymmetric simple exclusion process (9) in a single channel
or road; the exact correspondence indicates identical critical behavior in both
	the cases. The same universality for other cases of $\rho (i)$ in 
FBM suggests similar behavior for other equivalent ASEP cases as well. 

\section*{Acknowledgment:}
I am grateful to S.~M. Bhattacharjee, A. Chatterjee, A. Das,  P.~K. Mohanty
and R. B. Stinchcombe for useful discussions.

\end{document}